\documentclass{article}
\usepackage{lajolla2006}
\usepackage{graphicx,amsmath,amssymb,hhline}
\def\er{&\ensuremath{\pm}}

\def\ml#1{\multicolumn{2}{c|}{#1}}
\def\mc#1{\multicolumn{2}{c|}{#1}}
\def\mr#1{\multicolumn{2}{c||}{#1}}

\frompage{000} \topage{000}                                              
\def\rightmark{Universal scaling \ldots}
\def\leftmark{M. Csan\'ad, T. Cs\"org\H{o}, R. A. Lacey and B. L\"orstad}

\title{Universal scaling of the elliptic flow at RHIC}

\authors{
{M. Csan\'ad$^{1,2}$, T. Cs\"{o}rg\H{o}$^{3}$, R. A. Lacey$^{2}$ and
B. L\"{o}rstad$^{3}$ }\\[2.812mm]
{\normalsize \hspace*{-8pt}$^1$Department of Atomic Physics, ELTE,
Budapest,
P\'azm\'any P. 1/A, H-1117\\[0.2ex]
\hspace*{-8pt}$^2$ Department of Chemistry, SUNY Stony Brook, NY, 11794-3400, USA\\[0.2ex]
\hspace*{-8pt}$^3$ MTA KFKI RMKI, H - 1525 Budapest 114, P.O.Box
49, Hungary\\[0.2ex]
\hspace*{-8pt}$^3$ Div. Experimental High Energy Physics,
    Dept. Physics, Lund University,\\
    SE - 221 00 Lund, P.O.Box 118, Sweden \\[0.2ex]
}}

\abstract{
Recent PHOBOS measurements of the excitation function
for the pseudo-rapidity dependence of elliptic flow in Au+Au
collisions at RHIC, have posed a significant theoretical
challenge. We show here that these data
are described by the Buda-Lund model.
A universal scaling curve,  predicted  by the Buda-Lund hydro model,
describes not only PHOBOS data,
but also recent, detailed  PHENIX and STAR  elliptic flow data.
This is a consequence of perfect fluid hydrodynamics as implemented
in the Buda-Lund hydro model.
}

\keyword{elliptic flow, perfect hydrodynamics}

\PACS{25.75.Ld,24.10.Jv,24.10.Nz}

\begin{document}

\maketitle
\setcounter{page}{1}

\noindent {\bf Introduction, terminology.} One of the unexpected
results from experiments at the Relativistic Heavy Ion Collider
(RHIC) is the relatively strong second harmonic moment of the
transverse momentum distribution, referred to as the elliptic
flow. Measurements of the elliptic flow by the PHENIX, PHOBOS and
STAR collaborations (see
refs.~\cite{Back:2004zg,Adler:2003kt,Adams:2004bi,Adler:2001nb,Sorensen:2003wi})
reveal rich details in terms of its dependence on particle type,
transverse ($p_T$) and longitudinal momentum ($\eta$) variables,
and on the centrality and the bombarding energy of the collision.
In the soft transverse momentum region ($p_T \lesssim 2$\,GeV/c),
the measurements at mid-rapidity are found to be reasonably well
described by hydrodynamical
models~\cite{Adcox:2004mh,Adams:2005dq}. By contrast, differential
measurement of the pseudo-rapidity dependence of elliptic flow
$v_2(\eta)$, and its excitation function have resisted several
attempts at a description in terms of hydrodynamical models (but
see their new description by the SPHERIO
model~\cite{Grassi:2005pm}). Here we show that these data are
consistent with the theoretical and analytic predictions that are
based on Eqs. (1-6) of Ref.~\cite{hidro3}, that is, on perfect
fluid hydrodynamics.

    There is a confusion in the literature if perfect or perhaps
    an ideal fluid has been produced in Au+Au collisions at RHIC.
    The BNL press release~\cite{perfectliquid} uses the terminology
    ``perfect liquid". On the other hand,
    some other authors, for example
    refs.~\cite{Peshier:2005pp,Hirano:2005xf}
    are using a different terminology: ``ideal fluid".

    Although in many cases these different terms refer to
    the same set of equations, let us clarify that in fact
    there is a distinction between ideal and perfect liquids and the
    corresponding hydrodynamical equations.
    The expression ``ideal fluid" is frequently used either
    1: for an inviscid, incompressible fluid, or 2: for an inviscid
    fluid.
    These definitions of ``ideal fluid" are inequivalent,
    and one should note that definition 1 that includes the
    incompressibility is more frequent, as it is part of
    the primary physics
    courses: ``an ideal fluid keeps its volume but conforms to the
    outline of its container".

    Let us also consider the definitions of perfect liquids.
    A perfect liquid is  defined also in two different ways:
        1: A fluid which looks isotropic in its rest frame is called
    a perfect fluid.
    2: A fluid which has no shear stresses,
    viscosity or heat conduction is a perfect fluid.
    It turns out that these two definitions of the perfect fluids
    are equivalent.
    Perfect fluids are frequently used in general
    relativity to model idealized distributions of matter.
    This terminology, with clear definition of the energy-momentum
    tensor that is diagonal in the rest frame of the matter,
    has also been adopted by the Particle Data Group~\cite{pdg-perfectfluid}.
    When considering these questions on terminology,
    it is important to note that lack of viscosity
    does not yet imply that a fluid is perfect, as
    the lack of heat conduction and the vanishing of
    shear tensor needs also be discussed. Second,
    it is clear that we are discussing the time evolution
    of an exploding fireball, so an ideal fluid of a time independent
    volume is irrelevant for this problem,
    in contrast to def. 1 of ideal fluids.

    In conclusion, we will continue to use the well defined and
    unambigous terminology of ``perfect fluid" hydrodynamics in
    context of RHIC physics, and strongly recommend it also
    for other authors to avoid confusion on this subject.

\noindent {\bf The Buda-Lund hydro model.} The tool used to
describe the pseudorapidity-dependent elliptic flow is the
Buda-Lund hydro model. This hydro
model~\cite{Csorgo:1995bi,Csanad:2003qa} is based on exact
analytic hydro solutions discussed in
Refs.~\cite{Csorgo:2001xm,Csorgo:2003rt}. Buda-Lund type of exact,
parametric hydro solutions are being explored at the moment as
well. General, non-relativistic, ellipsoidally expanding
solutions~\cite{Csorgo:2001xm}, and relativistic but
non-accelerating solutions~\cite{Csorgo:2003rt,Sinyukov:2004am}
are known. An important missing link is the family of finite,
relativistic, accelerating solutions which currently being
pursued~\cite{Marci}.

The Buda-Lund hydro model is successful in describing the BRAHMS,
PHENIX, PHOBOS and STAR data on identified single particle spectra
and the transverse mass dependent Bose-Einstein or HBT radii as
well as the pseudorapidity distribution of charged particles in Au
+ Au collisions both at $\sqrt{s_{\rm{NN}}} =
130\,$GeV~\cite{ster-ismd03} and at $\sqrt{s_{\rm{NN}}} =
200\,$GeV~\cite{mate-warsaw03}. The model is defined with the help
of its emission function; to take into account the effects of
long-lived resonances, it utilizes the core-halo
model~\cite{Csorgo:1994in}, and characterizes the system with a
hydrodynamically evolving core and a halo of the decay products of
the long-lived resonances.

The elliptic flow would be zero in an axially symmetric case, so
we developed an ellipsoidal generalization of the model that
describes an expanding ellipsoid with principal axes $X$, $Y$ and
$Z$. Their derivatives with respect to proper-time (expansion
rates) are denoted by $\dot X$, $\dot Y$ and  $\dot Z$. The
generalization goes back to the original one, if the transverse
directed principal axes of the ellipsoid are equal, ie $X=Y$ (and
also $\dot X=\dot Y$). The deviation from axial symmetry can be
measured by the momentum-space eccentricity,
\begin{equation}
\epsilon_p = \frac{\dot X^2 - \dot Y^2}{\dot X^2 + \dot Y^2}.
\end{equation}
The exact analytic solutions of
hydrodynamics~\cite{hidro3,hidro1,hidro2}, which form the basis of
the Buda-Lund hydro model, develop Hubble-flow for late times, ie
$X \rightarrow_{\tau \rightarrow \infty} \dot X \tau$, so the
momentum-space eccentricity $\epsilon_p$ nearly equals space-time
eccentricity $\epsilon$. It turns out that the data fitting is
easier if instead of $\dot X$ and $\dot Y$ the eccentricity
variable $\epsilon $ is introduced and complemented by the average
radial flow defined as $\langle u_t \rangle = 1/2/\sqrt{1/\dot X^2
+ 1/\dot Y^2}$. Obviously, the expansion rates are given as and
$\dot X = \langle u_t \rangle / \sqrt{1 -  \epsilon_p}$ and $\dot
Y = \langle u_t \rangle / \sqrt{1 +  \epsilon_p}$.

Let us introduce $\Delta\,\eta$
additionally. It represents the elongation of the source expressed
in units of space-time rapidity. Let us consider furthermore that
at the freeze-out $\tau\,\Delta \eta = Z$ and $Z \approx \dot
Z\,\tau$, and so $\Delta \eta \approx \dot Z$. Hence, in this
paper we extract space-time eccentricity ($\epsilon$), average
transverse flow ($u_t$) and longitudinal elongation ($\Delta
\eta$) from the data, instead of $\dot X$, $\dot Y$ and $\dot Z$.

Although the proof of Ref.~\cite{hidro4} was given only for the
case of exact parametric solutions of non-relativistic
hydrodynamics, we expect a similar behavior at the end of
relativistic expansions just as well, because the acceleration of
the scales $(X,Y,Z)$ is due to pressure gradients. However, the
pressure hence its gradients as well tend to vanishing values in
the late stages of both the
non-relativistic~\cite{hidro1,hidro2,hidro4} and relativistic
\cite{hidro3,Csorgo:2003rt} exact parametric solutions of
hydrodynamics.

In the time dependent hydrodynamical solutions, these values
evolve in time, however, it was show in Ref.~\cite{hidro4} that
$\dot X$, $\dot Y$ and $\dot Z$, and so $\epsilon$, $u_t$ and
$\Delta \eta$ become constants of the motion in the late stages of
the expansion.

\noindent {\bf Elliptic flow from the model.} The result for the
elliptic flow is the following simple universal scaling law:
\begin{equation}
v_2=\frac{I_1(w)}{I_0(w)},
\label{e:v2w}
\end{equation}
where $I_n$ is the modified Bessel function of order  $n$.
For other flow coefficients, the prediction is also simple
and universal:
\begin{eqnarray}
v_{2n} &=& I_n(w)/I_0(w),\\
 v_{2n+1} &=& 0.
\end{eqnarray}

The model thus predicts a \emph{scaling law},
valid under certain conditions
detailed in Ref.~\cite{Csanad:2003qa}, fulfilled by these data :
every $v_2$ measurement falls on the same \emph{universal}
curve  of $I_1/I_0$ when plotted against $w$.

This means, that $v_2$ depends on any measurable quantity
only through the scaling parameter $w$. In practice this
implies that the detailed and experimentally known dependence of
the elliptic flow on transverse or longitudinal momentum,
center of mass energy, centrality, type of the colliding nucleus
etc is only apparent.

Before giving the definition of the scaling variable $w$
it is educational to introduce the effective temperatures
or slope parameters in the impact parameter direction ($T_x$)
and in the direction out of the reaction plane ($T_y$),
as follows:
\begin{eqnarray}
   T_{x}&=&T_0+\overline{m}_t \, \dot X^2
       \frac{T_0}{T_0 +\overline{m}_t a^2},\\
   T_{y}&=&T_0+\overline{m}_t \, \dot Y^2
     \frac{T_0}{T_0 +\overline{m}_t a^2},
\end{eqnarray}
  and
\begin{equation}
    \overline{m}_t = m_t \cosh(\eta_{s}-y).\label{e:mtbar}
\end{equation}
Here
$m_t = \sqrt{m^2 + p_x^2 + p_y^2}$ is the transverse mass,
$a=\langle \Delta T/T \rangle_t = (T_0 - T_s)/T_s$
measures the temperature gradient in the transverse direction,
at the mean freeze-out time, where
$T_0$ the central temperature at the mean
freeze-out time while $T_s$ is the temperature of the surface
of the fireball. Also we introduce a lengthscale, $R_s$,
defined by the transverse coordinate where the temperature
distribution decreases to half, as compared to the center,
so $T(r_x = r_y = R_s) = T_0/2$. In this notation, a flat temperature
profile corresponds to the $R_s \rightarrow \infty$ limit.

Furhermore, $\eta_{s}$ is the space-time rapidity of the
saddle-point (point of maximal emittivity), which depends on the
rapidity, the longitudinal expansion, the transverse mass and on
the central freeze-out temperature $\eta_{s}-y = \frac{y}{1+\Delta
\eta \frac{m_t}{T_0}},$ where $y = 0.5 \log(\frac{E + p_z}{E -
p_z})$ is the rapidity. Because of this, at midrapidity, where
PHENIX and STAR $v_2(p_t)$ data was taken, Eq.~(\ref{e:mtbar})
simplifies to $\overline{m}_t = m_t$. These definitions imply that
the effective temperature is transverse mass and rapidity
dependent both in the reaction plane and out of the reaction plane
direction,

The effective slope parameter, $T_*$,
of the asimuthally averaged single particle
spectra  can be defined as

\begin{equation}
\frac{1}{T_*} =
\frac{1}{2}
\left(
\frac{1}{T_x}+
\frac{1}{T_y}
\right) .
\end{equation}

A relativistic generalization of the kinetic energy, relevant for
the elliptic flow analysis is defined as
\begin{equation}
E_K = \frac{p_t^2}{2 \overline{m}_t}
\end{equation}

Finally, the eccentricity of the single particle spectra
can be defined as
\begin{equation}
\varepsilon = \frac{T_x - T_y}{T_x + T_y} .
\end{equation}

The {\it universal scaling variable} $w$ turns out to be the
product
\begin{equation}
    w  =
\frac{E_K}{2 T_*} \varepsilon
\end{equation}
It is interesting to note, that each of the kinetic energy term,
the effective temperature $T_*$ and the momentum space
eccentricity $\varepsilon$ are transverse mass and rapidity
dependent factors. However, for $\overline{m}_t / T_0 \gg 1$,
$\varepsilon \rightarrow \varepsilon_p =\frac{\dot X^2 -\dot
Y^2}{\dot X^2 +\dot Y^2}$ becomes independent of transverse mass
and rapidity.

This structure of the universal scaling variable suggests that the
detailed centrality, transverse mass and rapidity dependence of
the elliptic flow is only apparent, and the Buda-Lund model
predicts that when plotted against the scaling variable $w$, a
data collapsing behavior sets in and a universal scaling curve
emerges, which coincides with the ratio of the above two Bessel
functions. Additional details about the ellipsoidally symmetric
model and its result on $v_2(\eta)$ can be found in
Ref.~\cite{Csanad:2003qa}.

Eq.~(\ref{e:v2w}) depends, for a given centrality class, on
rapidity $y$ and transverse mass $m_t$. Before comparing our
result to the $v_2(\eta)$ data of PHOBOS, we thus performed a
saddle point integration in the transverse momentum variable and
performed a change of variables to the pseudo-rapidity $\eta=0.5
\log(\frac{|p| + p_z}{|p| - p_z})$, similarly to
Ref.~\cite{Kharzeev:2001gp}. This way, we have evaluated the
single-particle invariant spectra in terms of the variables $\eta$
and $\phi$, and calculated $v_2(\eta)$ from this distribution, a
procedure corresponding to the PHOBOS measurement described in
Ref.~\cite{Back:2004zg}.

Transverse momentum dependent elliptic flow data at mid-rapidity
can also be compared to the Buda-Lund result directly, as it was
done in e.g. Ref.~\cite{Csanad:2003qa}.

\noindent {\bf Comparison to experimental data.} When fitting only
$v_2(p_t)$ data of identified particles as measured by STAR and
PHENIX, we have found that fit parameters $u_t$ and $\epsilon$ are
essential. When analysing $v_2(\eta)$ data, we found that they can
be described by a two-parameter fit: parameter $\epsilon$ controls
the peak while parameter $\Delta \eta$ controlls the widht of
these distributions. Note that the same $\Delta\eta$ parameter
controlls the width of the pseudo-rapidity distribution as well,
but these simultaneous fits will be reported elsewhere. The
quality of our fits was found to be insensitive to the precise
value of other parameters of the model. Hence we have fixed their
values. We also excluded points with large rapidity from lower
center of mass energies $v_2(\eta)$ fits ($\eta>4$ for 19.6\,GeV,
$\eta>4.5$ for 62.4\,GeV) and points with large transverse
momentum ($p_t>1.3$\,GeV) from $v_2(p_t)$ fits.

\begin{figure}
\vspace{0.5truecm}
\begin{center}
  \includegraphics[height=100pt]{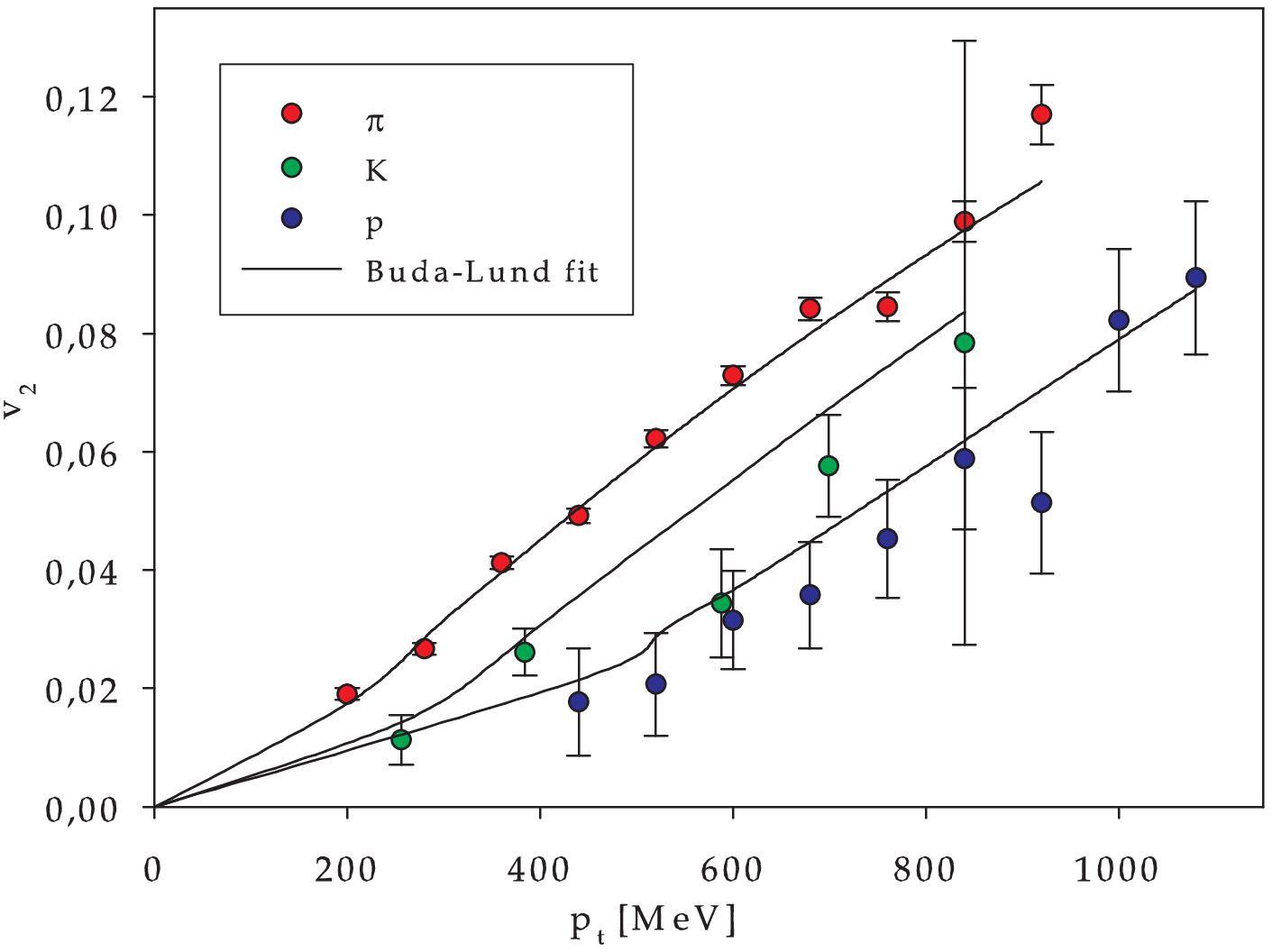}
  \includegraphics[height=100pt]{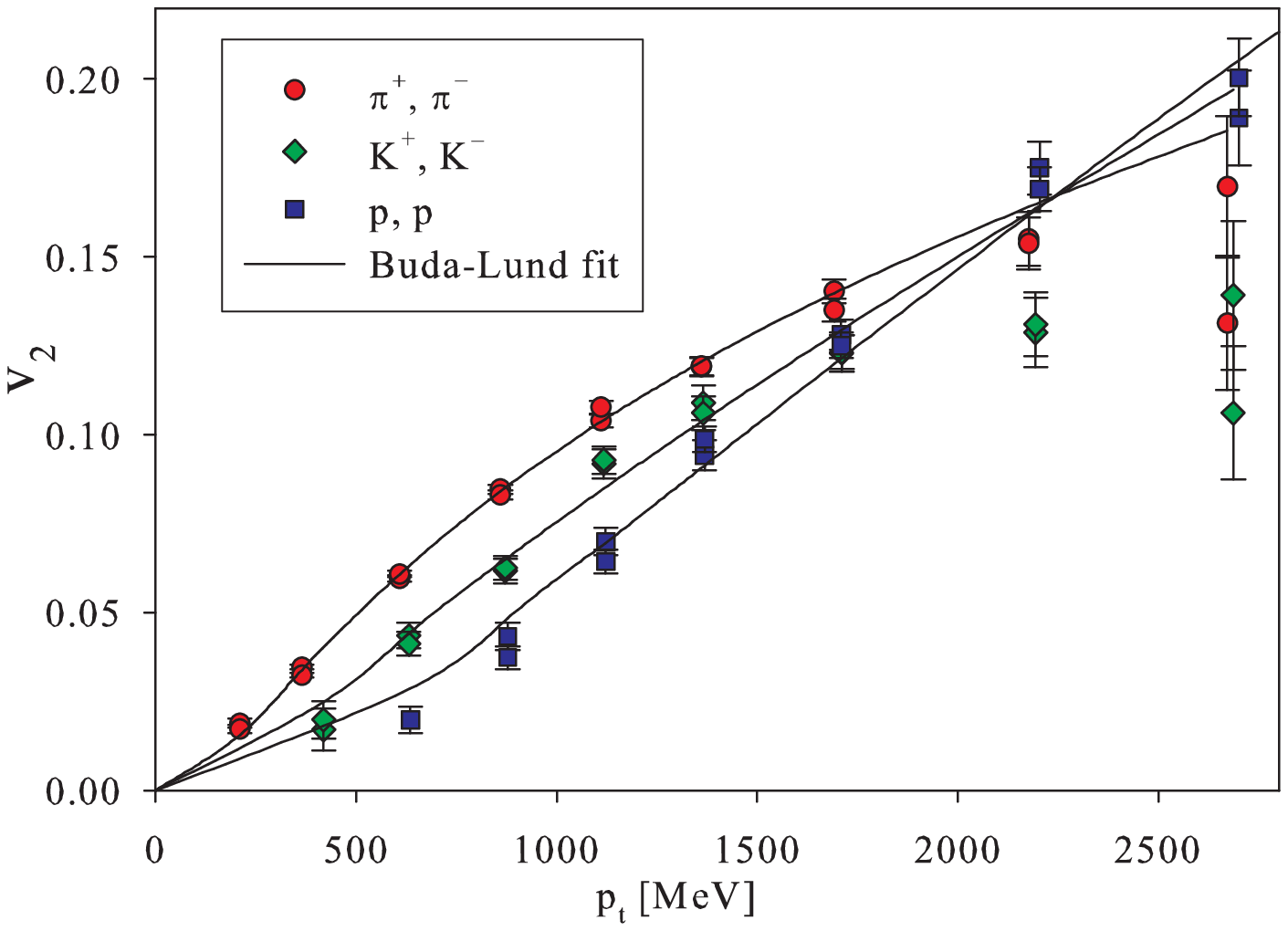}
\\[4ex]
\includegraphics[width=300pt]{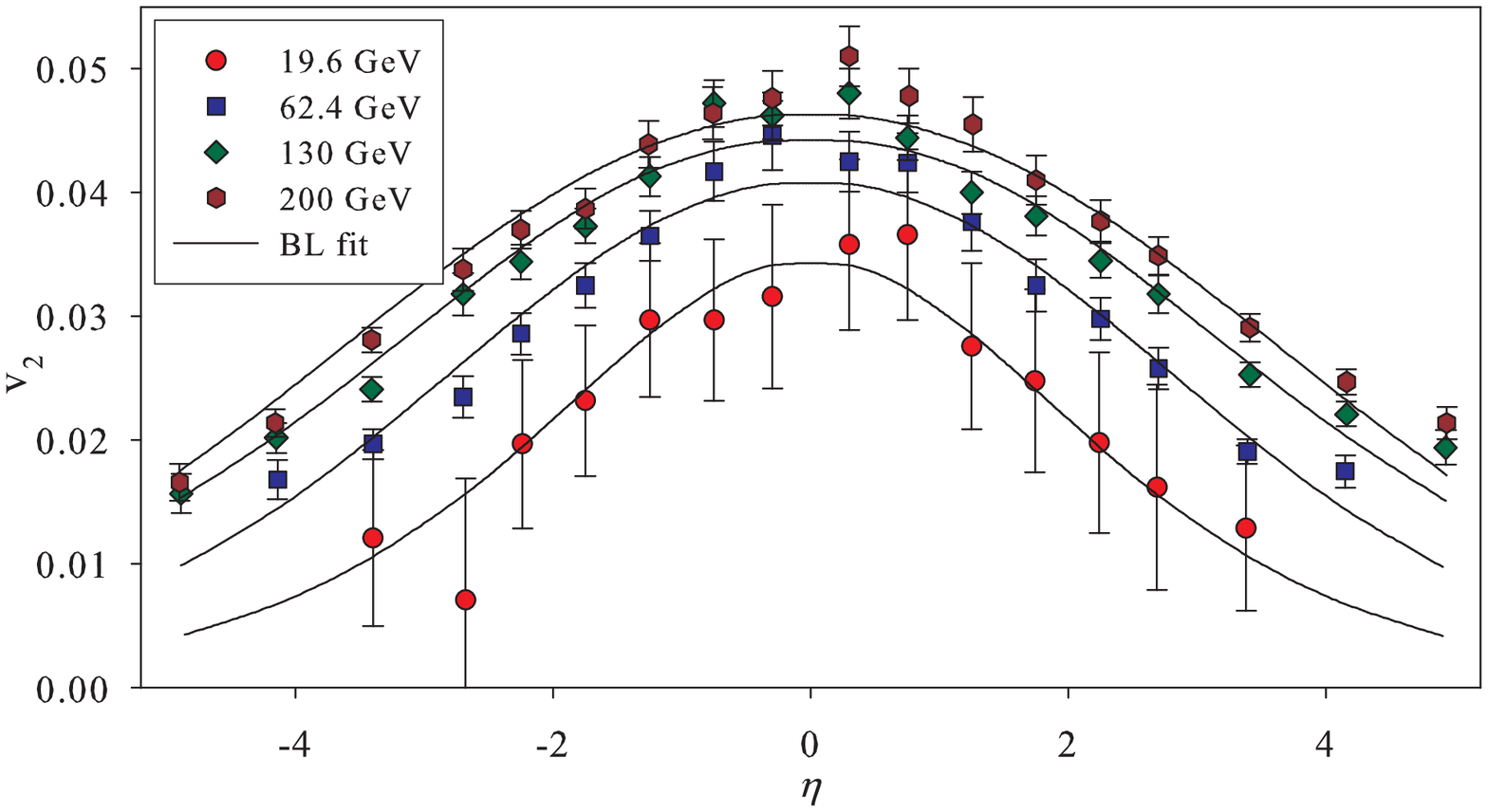}\\[4ex]
  \includegraphics[width=300pt]{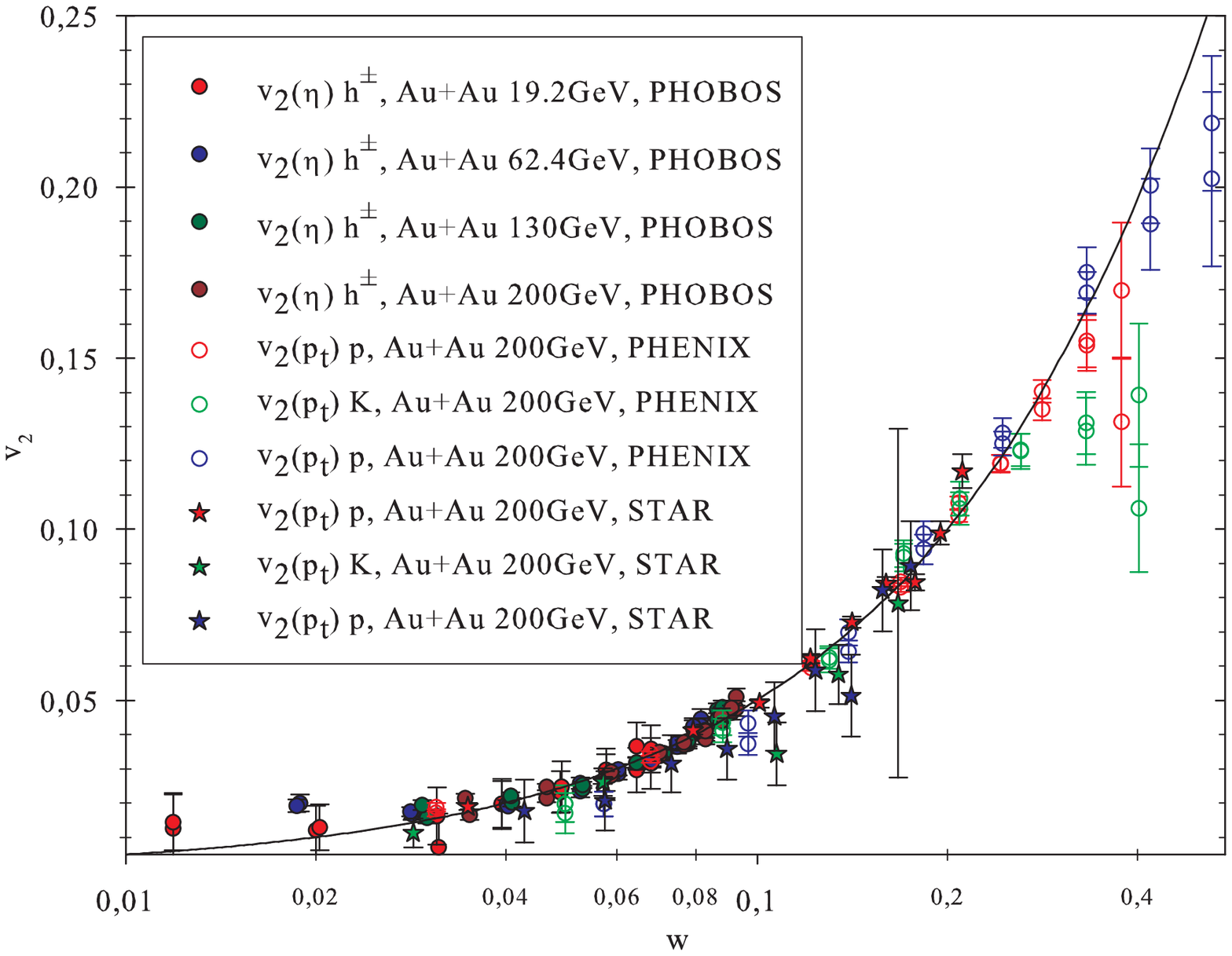}
\end{center}
  \caption{
STAR~\cite{Adams:2004bi} (top left), PHENIX~\cite{Adler:2003kt} (top right)
and PHOBOS~\cite{Back:2004zg} (middle) data on elliptic flow, $v_2$, plotted
versus $p_t$ and $\eta$ and fitted
   with Buda-Lund model.
   Bottom: elliptic flow versus variable $w$ is plotted.
  The data points show the  predicted~\cite{Csanad:2003qa} universal scaling.  }
  \label{f:v2fit}
\end{figure}

Fits to PHOBOS data of Ref.~\cite{Back:2004zg}, PHENIX data of
Ref.~\cite{Adler:2003kt} and STAR data of Ref.~\cite{Adams:2004bi}
are shown in the top three panels of Fig.~\ref{f:v2fit}. Bottom
panel of Fig.~\ref{f:v2fit} demonstrates that the investigated
PHOBOS, PHENIX and STAR data points follow the theoretically
predicted scaling law. The fitting package is available at
Ref.~\cite{blcvs}, and the fits can be redone on the interactive
online Buda-Lund page~\cite{blfit}.

    In contrast, Ref.~\cite{Hirano:2005xf} considered the description of the centrality
    and pseudorapidity dependence of the elliptic flow in
    200\,GeV Au+Au collisions as measured by the PHOBOS collaboration
    in Ref.~\cite{Back:2004zg}, however, this attempt failed to describe
    these data in a unified framework and concluded, that
    ``whether all of the observed discrepancies between elliptic flow
    data and \ldots\ simulations can be blamed on 'late
    hadronic viscosity' \ldots\ depends on presently unknown details of
    the initial state of the matter formed in heavy-ion
    collisions".

    This failure can be contrasted to the success of the Buda-Lund
    hydro model, that analytically predicted
    the scaling properties of elliptic flow in 2003.
    The particle mass, transverse momentum, pseudorapidity
    and centrality dependence of the elliptic
    flow at RHIC are being fully explored only
    now in experimental studies. We think that one of the important
    elements of success was that the Buda-Lund hydro model was based
    on exact, parametric and analytic solutions of
    hydrodynamics. These solutions were found recently
    for both non-relativistic and relativistic kinematic domain.
    Non-relativistic solutions include refs.~\cite{hidro1,hidro2,hidro4}.
    Relativistic solutions include not only the Hwa-Bjorken and the
    Hubble solutions but their new relativistic generalizations
    for 1+1 dimensional, 1+3 dimensional axially symmetric,
    and ellipsoidally symmetric solutions,
    see refs.~\cite{hidro3,Csorgo:2003rt}.

\noindent {\bf Conclusions.} In summary, we have shown that the
excitation function of the transverse momentum and pseudorapidity
dependence of the elliptic flow in Au+Au collisions is well
described by the Buda-Lund model. We have thus provided a
quantitative evidence of the validity of the perfect fluid picture
of soft particle production in Au+Au collisions at RHIC. This
perfect fluid extends far away from mid-rapidity. The universal
scaling of PHOBOS $v_2(\eta)$ and PHENIX and STAR $v_2(p_t)$,
expressed by Eq.~(\ref{e:v2w}) and illustrated by
Fig.~\ref{f:v2fit} provides a successful quantitative as well as
qualitative test for the appearance of a perfect fluid in Au+Au
collisions at various colliding energies at RHIC.

\noindent {\bf Acknowledgments.} T. Cs. is greatful for B. L. for
a kind hospitality at the University of Lund. This research was
supported by the NATO Collaborative Linkage Grant PST.CLG.980086,
by the Hungarian - US MTA OTKA NSF grant INT0089462, by the OTKA
grants T038406, T049466, by a KBN-OM Hungarian - Polish programme
in S\&T.

\begin{table}
\begin{tabular}{||c||rl|rl|rl|rl||}
\hhline{|t:=:t:========:t|}
           &\multicolumn{8}{c||}{PHOBOS Au+Au $v_2(\eta)$}\\
\hhline{||~||--------||}
           &\ml{19.6 GeV}&\mc{62.4 GeV}&\mc{130 GeV}&\mr{200 GeV}\\
\hhline{|:=::========:|}
$\epsilon$  &0.30\er0.03&0.37\er0.01&0.40\er0.01&0.42\er0.01\\
\hhline{||-||--------||}
$\Delta\eta$&2.0\er0.3  &2.28\er0.05&2.60\er0.04&2.70\er0.04\\
\hhline{|:=::========:|}
$\chi^2$/NDF&\ml{1.9/11}&\mc{21/13} &\mc{36/15} &\mr{26/15}\\
\hhline{||-||--------||}
conf. level &\ml{100\%} &\mc{19\%}  &\mc{0.78\%}&\mr{6.2\%} \\
\hhline{|b:=:b:========:b|}
\end{tabular}
\begin{tabular}{||c||rl|rl|rl||rl||}
\hhline{|t:=:t:======:t:==:t|}
           &\multicolumn{6}{c||}{PHENIX 200GeV Au+Au $v_2(p_t)$}&\multicolumn{2}{c||}{STAR}\\
\hhline{||~||------||--||}
           &\ml{$\pi$}    &\mc{K}     & \mr{p}      &\mr{$\pi$, K, p}\\
\hhline{|:=::======::==:|}
$\epsilon$  &0.31\er0.003 &0.32\er0.04&0.3\er0.1    &0.38\er0.01\\
\hhline{||-||------||--||}
$u_t$       &1.2\er0.1    &0.7\er0.12 &0.6\er0.3    &1.0\er0.1\\
\hhline{|:=::======::==:|}
$\chi^2$/NDF&\ml{21/7}    &\mc{5/5}   &\mr{10/2}    &\mr{45/21}\\
\hhline{||-||------||--||}
conf. level &\ml{2\%}     &\mc{80\%}  &\mr{8\%}     &\mr{0.5\%}\\
\hhline{|b:=:b:======:b:==:b|}
\end{tabular}
  \caption{Results of fits to data of PHOBOS~\cite{Back:2004zg}, PHENIX~\cite{Adler:2003kt} and STAR~\cite{Adams:2004bi}.
  Fixed parameters: $T_0=200$\,MeV, $a=$1,
and for $v_2(\eta)$, $u_t=1.5$. Note that data from PHOBOS, PHENIX
and STAR have different centrality selection. }\label{t:fit}
\end{table}

\vfill\eject
\end{document}